\documentclass[12pt]{article}

\usepackage[left=2.8cm,right=2.8cm,top=2.9cm,bottom=2.9cm]{geometry}
\usepackage[french,english]{babel}
\usepackage[colorlinks=true,linkcolor=black,citecolor=blue,urlcolor=blue,filecolor=black]{hyperref}

\usepackage{color,graphicx}

\usepackage[utf8]{inputenc}
\usepackage[T1]{fontenc}
\usepackage{latexsym}
\usepackage{amsmath}
\usepackage{slashed}
\usepackage{amsfonts}
\usepackage{verbatim}
\usepackage{mathrsfs}
\usepackage{dsfont}

\newcommand{\be}{\begin{eqnarray}}
\newcommand{\en}{\end{eqnarray}}
\newcommand{\badat}{\begin{alignedat}}
\newcommand{\eadat}{\end{alignedat}}
\newcommand{\bitm}{\begin{itemize}}
\newcommand{\eitm}{\end{itemize}}
\newcommand{\bmat}{\begin{pmatrix}}
\newcommand{\emat}{\end{pmatrix}}
\newcommand{\ba}{\begin{align}}
\newcommand{\bas}{\begin{align*}}
\newcommand{\ab}{\end{align}}
\newcommand{\bse}{\begin{subequations}}
\newcommand{\ese}{\end{subequations}}

\def\bea{\begin{eqnarray}}
\def\eea{\end{eqnarray}}
\def\ba{\begin{array}}
\def\ea{\end{array}}
\def\bec{\begin{center}}
\def\ec{\end{center}}
\def\ba{\begin{align}}
\def\ena{\end{align}}

\def\12{\frac{1}{2}}

\begin{document}

\title{\textbf{The BMS-like symmetry of extremal horizons}}

\author{Gaston Giribet}
\date{}
\maketitle

\begin{center}

\smallskip
\smallskip

{Departamento de F\'{i}sica, Universidad de Buenos Aires and IFIBA-CONICET}\\
{{\it Ciudad Universitaria, Pabell\'{o}n I, 1428, Buenos Aires, Argentina.}}

\smallskip
\smallskip

\end{center}

\bigskip

\begin{abstract}
I revisit the calculation of infinite-dimensional symmetries that emerge in the vicinity of isolated horizons. I focus the attention on extremal black holes, for which the isometry algebra that preserves a sensible set of asymptotic boundary conditions at the horizon strictly includes the BMS algebra. The conserved charges that correspond to this BMS sector, however, reduce to those of superrotation, generating only two copies of Witt algebra. For more general horizon isometries, in contrast, the charge algebra does include both Witt and supertranslations, being similar to BMS but {\it s.str.} differing from it. 
\[\]

This work has been prepared for the proceedings of the XXII Simposio Sofichi 2020, held in Chile in November 2020. The material herein is based on my work in collaboration with Laura Donnay, Hern\'an Gonz\'alez and Miguel Pino, and it is included in \cite{DGGP1, DGGP2}. 
\end{abstract}

\newpage

\section{Introduction}

The isometries of asymptotically flat spacetimes near null infinity $\mathcal{I}^{\pm }$ are generated by the Bondi-Metzner-Sachs (BMS) algebra \cite{BMS1, BMS2, BMS3}, which consists of a semidirect sum of the Lorentz group and an infinite-dimensional Abelian ideal known as supertranslations. This algebra and its extensions \cite{Barnich0} have recently been reconsidered from a modern perspective \cite{Barnich2, Barnich3, Barnich4, Barnich6}, and this led to investigate its relation to topics as diverse as holography in asymptotically flat spacetimes \cite{Barnich3}, soft graviton theorems \cite{Strominger1, Strominger2}, gravitational memory effects \cite{Strominger3}, the infrared structure of gauge theories \cite{Strominger}, and black hole physics \cite{HPS1, HPS2}. 

In \cite{Hawking}, Hawking suggested that BMS type supertranslations should appear not only at $\mathcal{I}^{\pm }$ but also in the near horizon region of black holes \cite{HPS1, HPS2}. In \cite{DGGP1}, these symmetries were explicitly found, along with an infinite set of new superrotation currents that form two copies of the Witt algebra in semidirect sum with supertranslations. This was further investigated in \cite{DGGP2, DGGP3, Laura, Afshar} and references thereof. The near horizon symmetry found in \cite{DGGP1}, nevertheless, does not exactly coincide with BMS symmetry\footnote{BMS symmetry in the near horizon region has recently been found in \cite{Troncoso}.}: While both symmetry algebras share many properties and include supertranslations, they actually have different structure constants. 

Here, I will revisit the calculation of infinite-dimensional symmetries that emerge in the vicinity of isolated horizons \cite{DGGP1, DGGP2, DGGP3}, focusing my attention on extremal black holes. For the latter, we will see that the isometry algebra that preserves a sensible set of asymptotic boundary conditions at the horizon strictly includes the BMS algebra \cite{BMS1, BMS2, BMS3}. This will be discussed in section 3, after reviewing the near horizon symmetries in section 2.

\section{The infinite symmetries of black hole horizons}

Let us start by considering the near horizon geometry of stationary black holes. Close to the event horizon, we can always consider the spacetime metric in the form \cite{Booth, Moncrief}
\begin{equation}\label{ParaEmpezar}
ds^2= -2 \kappa\, \rho \, dv^2 +2\, d\rho dv + 2N_A\,\rho\, dz^Adv+ \Omega_{AB}\, dz^Adz^B + ...
\end{equation}
where the ellipsis stand for subleading terms. $z^A$ with $A=1,2$ represents coordinates on constant-$v$ slices of the horizon, $v\in \mathbb{R}$ being the advanced time (null) coordinate. $\rho \in \mathbb{R}_{\geq 0}$ measures the distance from the horizon, which is located at $\rho =0$. Functions $N_A$ and $\Omega_{AB}$ depend on the two coordinates $z^A$, and in principle they might depend on time as well\footnote{The isolated horizon condition later excludes the dependence on $v$.}. The specific form of the subleading terms is given by
\begin{equation}\label{RNmetricBdyCond}
g_{vv}=-2 \kappa \, \rho +\mathcal{O}(\rho^2)\,, \ \ \
g_{vA}=  N_A(v, z^B)\, \rho +\mathcal{O}(\rho^2)\,, \ \ \
g_{AB}=\Omega_{AB}(v, z^C)+\mathcal{O}(\rho)\,,
\end{equation} 
where $\mathcal{O}(\rho^n)$ stand for functions on the coordinates whose dependence with $\rho $ damps off at least as fast as $\sim \rho^n$ when $\rho $ tends to zero. $\kappa $ in (\ref{RNmetricBdyCond}) represents the surface gravity at the horizon, and so we will consider it to be constant, cf. \cite{DGGP2}. For the other components of the metric, we consider the gauge fixing conditions $g_{\rho \rho }=0,\, g_{v\rho }=1,\,  g_{A\rho }=0$.

Now, let us study the diffeomorphisms that preserve such a near horizon form of the metric. To do that, we compute the Lie derivative $\delta _{\xi }g_{\mu \nu}=\mathcal{L}_\xi g_{\mu \nu}$ with respect to a vector field $\xi=\xi^\mu \partial_\mu$, and demand it to preserve the functional form (\ref{ParaEmpezar}). More precisely, we require (\ref{ParaEmpezar}) with the expansion (\ref{RNmetricBdyCond}) to be preserved, but allow the specific functions $N_A$ and $\Omega_{AB}$ to change. The gauge fixing conditions being preserved implies $\mathcal{L}_\xi g_{\rho \rho}=0$, $\mathcal{L}_\xi g_{v \rho }=0$ and $\mathcal{L}_\xi g_{\rho A}=0$. All these requirements result in the following asymptotic Killing vectors
\begin{equation}
\badat{3}
\label{chisNonExt}
&\xi^{v} = T+ \mathcal{O}(\rho ), \\
&\xi^{\rho } =-\partial_v T\, \rho  + \frac{1}{2 }{\Omega}^{AB} N_A \, \partial_{B} T\, \rho^2 + \mathcal{O}(\rho^3),  \\
&\xi^{A} = L^A + {\Omega}^{AB}\, \partial_{B} T\, {\rho}  +  \mathcal{O}(\rho^2),
\eadat
\end{equation}
where $T$ is a function on $v$ and $z^A$, and $L^A$ is a function of $z^A$. The expressions can be considerably simplified if we consider the conformal gauge, 
which implies that $L^A$ are conformal Killing vectors on the 2-sphere: $\partial_{\bar z}L^{z}=\partial_{z}L^{\bar z}=0$, with $z,\, \bar{z}$ being holomorphic and anti-holomorphic coordinates on the spacelike sections of the horizon. Generalization to the full $\text{Diff}(S^2)$ symmetry is also possible, cf. \cite{Penna}. 

The variations of the functions $\kappa,\, N_A,\, \Omega_{AB}$ after the action of (\ref{chisNonExt}) are found to be
\begin{equation}
\badat{5} \label{deltaRN}
&\delta_{\xi} \kappa = \kappa \partial_v T + \partial_v^2 T\,,  \\ 
&\delta_{\xi } N_A =\mathcal{L}_L N_{A}+ T \partial_vN_A -2\kappa  \partial_A T -2\partial_v\partial_A T +\Omega^{BC} \partial_v\Omega_{AB} \partial_C T \,,\\
&\delta_{\xi }  \Omega_{AB} = T\partial_v\Omega_{AB} + {\mathcal L}_{L}\Omega_{AB}\,.
\eadat
\end{equation}

At this point, we may assume functions $N_A$ and $\Omega_{AB}$ to depend only on $z^A$. This follows from the isolated horizon condition together with integrability of the Noether charges \cite{DGGP2}. In addition, we impose the condition $\delta_{\xi} \kappa =0$; that is to say, we consider functional variations that keep the surface gravity unchanged. 
\begin{figure}[ht!]
\begin{center}
\includegraphics[width=5in]{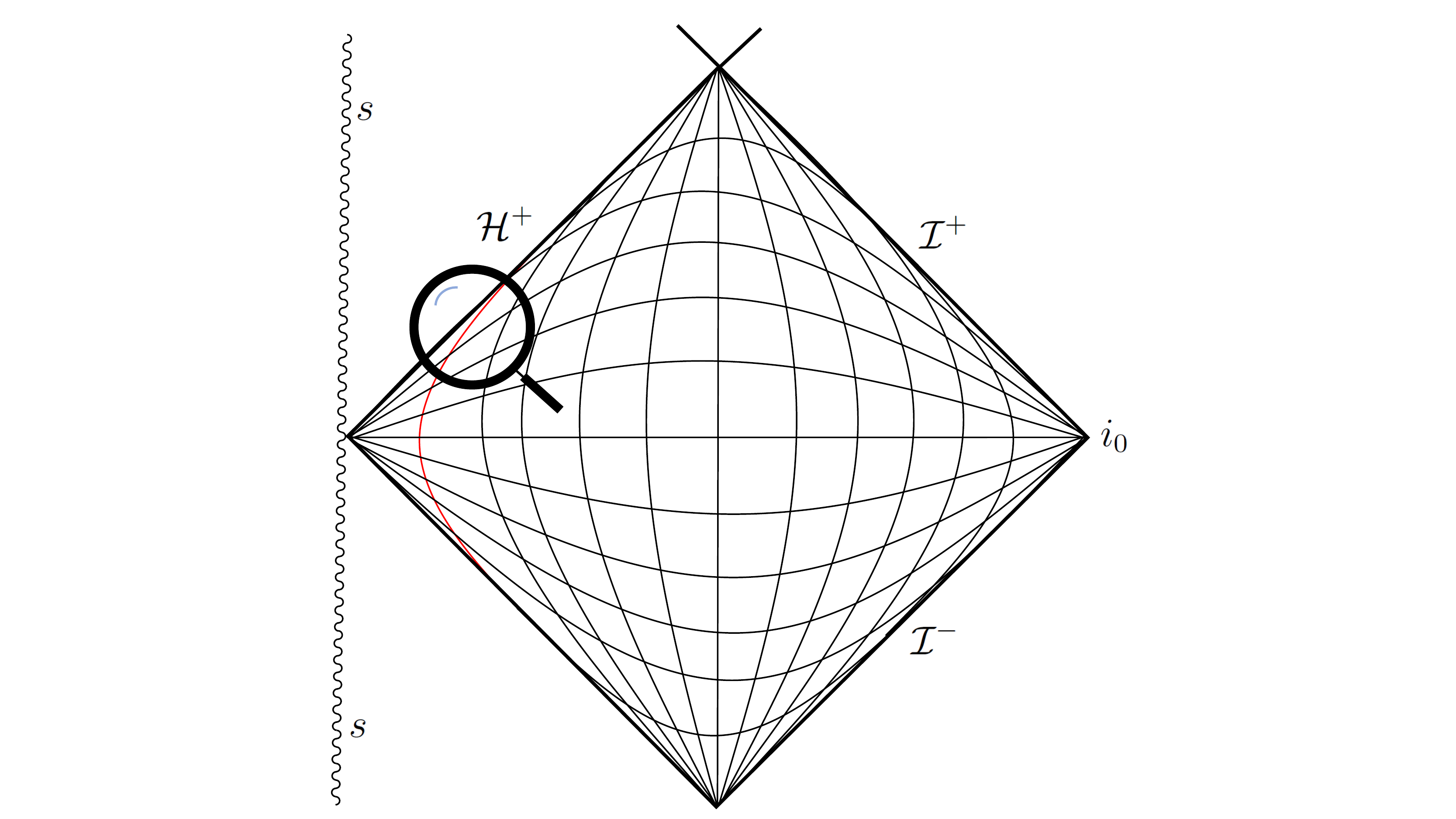}
\caption{Penrose diagram of an extremal, asymptotically flat black hole. $\mathcal{I}^+ $ ($\mathcal{I}^-$) is the null future (past) infinity, $\mathcal{H}^+$ is the future horizon, ${i}_0 $ is spatial infinity, and $s$ stands for the timelike singularity. The red line delimits the near horizon region.}
\label{Figura}
\end{center}
\end{figure}

Variations (\ref{deltaRN}) generate a Lie algebra realized by the application 
\begin{equation}
[\delta _{\xi_1 },\delta _{ \xi_2 }]g_{\mu \nu}=\delta _{\hat \xi }g_{\mu \nu}.\label{esto}
\end{equation}
Since the asymptotic Killing vector $\xi$ defined in (\ref{chisNonExt}) depends not only on the spacetime coordinates but also on the fields, the Lie product that defines (\ref{esto}) is not given by the usual product $[\xi_1,\xi_2]$ but rather by the modified Lie bracket \cite{Barnich01}; namely
\begin{equation}
\hat \xi=[\xi_1,\xi_2]+\delta_{\xi2}\xi_1-\delta_{\xi1}\xi_2\, ,
\end{equation}
which suffices to take into account the dependence of $\xi $ on the metric functions. In this way, we find the following algebra of diffeomorphisms \cite{DGGP1}
\begin{equation} \label{modifiedLie}
\badat{3}
&\hat T=T_1 \partial_{v} T_2+L_1^A \partial_{A} T_2-
        T_2 \partial_{v} T_1-L_2^A \partial_{A} T_1\,,\\
& \hat L^A=L_1^B  \partial_{B}  L_2^A - L_2^B  \partial_{B}  L_1^A\,.
\eadat
\end{equation}
$T$ and $L^A$ being functions of $z^A$, this encodes an infinite-dimensional algebra. It consists of two copies of the Witt algebra, generated by $L^z(z)$ and $L^{\bar{z}}(\bar{z})$, in semi-direct sum with the current algebra generated by $T(z,\bar{z})$. It is worth noticing that, despite some similarities, this algebra is different from the BMS algebra. The structure constants of both algebras differ, although they share a set of subalgebras, including an infinite-dimensional Abelian ideal. In the next section we will show that in the case of extremal black holes, a special set of near horizon isometries generating the full BMS algebra can be defined. 

\section{BMS isometries of extremal horizons}

Now, we consider extremal horizons, for which $\kappa =0$. In that case, equations (\ref{deltaRN}) together with the condition $\delta_{\xi }\kappa =0$ imply $\partial_v^2 T=0$, yielding the general solution
\begin{equation}\label{fExt}
T=P(z,\bar{z})+v\ J(z,\bar{z}) .  
\end{equation} 
This contains a linearly growing term in the advanced time, so differing from the non-extremal case for which an exponential dependence is found. In terms of these functions, algebra (\ref{modifiedLie}) reads
\begin{equation}\label{Esto}
\badat{3}
&\hat P=P_1 J_2+L_1^A \partial_{A} P_2-
        P_2 J_1-L_2^A \partial_{A} P_1,\\
&\hat J=L_1^A  \partial_{A}  J_2 -
L_2^A  \partial_{A}  J_1\,, \\
& \hat L^A=L_1^B  \partial_{B}  L_2^A - L_2^B  \partial_{B}  L_1^A\,,
\eadat
\end{equation}
which, apart from the ubiquitous Witt algebras, contains two supertranslation currents -- one of them actually resembling a superdilation operation--. The Noether charges associated to~\eqref{Esto}, however, only depend on the currents $J(z, \bar z) $, $L^{z }(z)$ and $L^{\bar z }(\bar z)$, while, in contrast to the case of non-extremal horizons, do not involve the time-independent part $P(z,\bar z )$. These Noether charges can be computed using the Barnich-Brandt formalism \cite{BarnichBrandt}; they turn out to be finite, integrable and conserved \cite{DGGP2}, and take the relatively simple form 
\begin{equation}\label{QhorExtRN}
Q[J,L^{A}]=\frac{1}{16\pi G} \int d^2z \, \sqrt{ \det \Omega_{AB} }\,   \Big( 2J - L^A N_A \Big) .
\end{equation}
The zero modes of these charges reproduce the Wald entropy (for $J=2\pi $) and the angular momentum (for $L^z-L^{\bar z}=1$); see \cite{DGGP2} for details.

What we want to argue here is that the set of transformations (\ref{Esto}) strictly includes the BMS algebra. To see this, consider the particular transformations obeying
\begin{equation}
J=\frac 12 D_AL^A\,.\label{Magic}
\end{equation}
This particular subset of isometries yields the algebra\footnote{I thank Laura Donnay for explaining this to me.}
\begin{equation}\label{BMS}
\badat{3}
&\hat L^A=L_1^B  \partial_{B}  L_2^A - L_2^B  \partial_{B}  L_1^A\,\\
&\hat P = \frac 12 P_1D_AL_2^A + L_1^AD_A P_2 
         -\frac 12 P_2D_AL_1^A - L_2^AD_A P_1 \,,
\eadat
\end{equation}
which is actually the BMS algebra \cite{BMS1, BMS2, BMS3} augmented with superrotations \cite{Barnich2, Barnich3, Barnich4, Barnich6}. While function $P(z,\bar z )$ represents supertranslations, functions $L^z(z)$ and $L^{\bar{z}}(\bar z)$ satisfy two copies of the Witt algebra in semi-direct sum with supertranslations. It is worth noticing that, while (\ref{modifiedLie}) already realizes the semidirect sum of two Witt algebras and a current algebra, the structure constants in (\ref{BMS}) are different and do reproduce those of \cite{Barnich3}; see Eq. (4.11) therein. A link between suertranslations and superrotations like the one expressed by (\ref{Magic}) also appears in the construction of \cite{Felipe}.

Nevertheless, it turns out that this BMS form of the asymptotic isometries algebra is somehow {\it virtual}, as the algebra of charges does not exhibit such a form: After imposing (\ref{Magic}), the contribution of the supertranslation $J(z,\bar{z})$ to the charges vanishes since the first term in the integrand of (\ref{QhorExtRN}) becomes a total derivative. Therefore, the charges in that case reduce to those of superrotations, {\it i.e.} two copies of Virasoro algebra with vanishing central charge.

In the recent work \cite{Troncoso}, a realization of BMS algebra in the near horizon of non-extremal black holes has been found, along with many other algebraic structures depending on the boundary conditions. It would be interesting to understand in detail the relation with that work and with other recent works that consider infinite-dimensional horizon symmetries. For the particular case of extremal horizons, it would be interesting to investigate the connection with Kerr/CFT correspondence \cite{KerrCFT}. This relates to the question as to whether there exists a natural way of defining horizon boundary conditions that lead to a central extension of the symmetries found in \cite{DGGP1, DGGP2}, likely involving a different kind of departure from the horizon, more drastic in a sense.

\subsection*{Acknowledgments}

I am grateful to Laura Donnay, Hern\'an Gonz\'alez, Julio Oliva, Miguel Pino, Andrea Puhm, and Felipe Rosso for collaborations on related subjects. I also thank the organizers of the XXII Simposio SOFICHI 2020. This work has been partially supported by CONICET through the grant PIP 1109-2017.






\end{document}